# Characterization of Morphology Evolution in a Polymer-Clay Nanocomposite using Multiscale Simulations

*Parvez Khan[a,b§], Ankit Patidar[a§], Gaurav Goel[a*]*

[a]Department of Chemical Engineering, Indian Institute of Technology Delhi, New Delhi 110016, India

[b]Department of Chemical Engineering, Aligarh Muslim University, Aligarh, India

* Email: goelg@chemical.iitd.ac.in

[§] Parvez Khan and A. Patidar contributed equally to this work

ABSTRACT

Molecular simulations provide an effective route for investigating morphology evolution and structure-property relationship in polymer-clay nanocomposites (PCNCs) incorporating layered silicates like montmorillonite (MMT), an important class of materials that show a significant enhancement over the constituent polymer for several properties. However, long relaxation times and large system size requirements limit the application to systems of practical interest. In this work, we have developed a coarse-grained (CG) model of organically modified MMT (oMMT) compatible with the MARTINI force field, a chemically-specific interaction model with high transferability. The dispersive and polar components of cleavage energy, basal spacing, and mechanical properties of MMT with tetramethylammonium (TMA) as intergallery ions were used to obtain a rational estimate for clay particle MARTINI bead types in accordance with the polarity of the functional group. The CG model provided accurate concurrent estimates for the structural, thermodynamic, and dynamical properties of PE in a PE/TMA-MMT PCNC, with less than 4% deviation from all atom (AA) simulations. The slow clay-induced redistribution of the PE-b-PEG block copolymer in the PCNCs was investigated using the developed CG model, with conformational changes occurring over a microsecond timescale. The preferential interaction coefficient and cluster analysis of individual blocks of PE-b-PEG were used to study the effect of clay arrangement (exfoliated versus tactoid) on copolymer reorientation and assembly at the clay surface. We find that the oMMT coated with PE-b-PEG acts as a neutral surface (small difference in polymer-polymer and polymer-oMMT+PE-b-PEG enthalpic interactions) and the primary influence of the nanofiller is a result of confinement and steric effect of the clay sheets on the PE chains. Finally, several different PCNC morphologies obtained from long CG simulations were backmapped to AA resolution for accurate calculation of mechanical and physical properties. This work offers a computationally efficient multiscale simulation framework for the accurate determination of the morphology and mechanical performance of PCNCs, enabling rational material design.

# 1. Introduction

The design and optimization of polymer nanocomposites (PNCs) is of high functional and technological importance. These hybrid materials, consisting of polymer matrices with nanoscale fillers, exhibit a range of novel properties that can be tailored to meet diverse applications, from advanced electronics to aerospace engineering.[1–10] These applications benefit from a significant improvement in one or more of the thermal, mechanical, electrical, optical, and barrier properties compared to the unfilled polymer.[11–15,15–19] The properties of these hybrid materials are intricately linked to the overall morphology of the composite[20,21] and therefore, have received significant attention in both fundamental and applied sciences.[22,23]

Spherical metallic or latex nanoparticles, carbon nanotubes, sheet/platelet nanoparticles such as graphene sheets and phyllosilicates are some of the commonly used nanofillers. PNCs consisting of highly anisotropic layered-silicate (clay) nanoparticles,[24] referred to as polymer-clay nanocomposites (PCNCs), provide a distinct advantage because of their large surface-to-volume ratio, leading to a significant effect on mechanical, structural, and barrier properties even at low loadings.[25–27] For example, Nylon 6-clay nanocomposite exhibited superior mechanical (~ 68% increment in Young's modulus) and thermal properties (~ 75% increase in heat distortion temperature) on the addition of 4-6 % w/w montmorillonite (MMT), an aluminosilicate clay.[28] The enhancement in PCNC properties directly depends on the extent of dispersion of clay tactoids in the polymer matrix,[29,30,31] which is primarily affected by the chemical makeup of the polymer and the clay, with specifics of processing (kinetics) having a negligible effect. Mg substitution (in octahedral or tetrahedral layer) in aluminosilicate clays leads to a net charge and provides a route for surface modification of the clay, wherein, the interlayer cation can either be metallic (e.g., in Na-MMT) or an alkylammonium ion containing one or more C5-C18 alkyl units[32] (referred to as organically-modified MMT, OM-MMT). A polymer matrix with fully exfoliated cloisite 15A (an OM-MMT) exhibited a larger improvement in Young's modulus (49.13%) of thermoplastic starch (TPS) films than those with intercalated or tactoid assemblies.[33] Similarly, the addition of PE-g-MA to improve the dispersion of Na-MMT in polyethylene (PE) led to a 30% increase (w.r.t PE) in the Young's modulus.[34]

In general, clay dispersion is dependent on a combination of several parameters, such as aspect ratio, composition (concentration of clay and polymer), molecular weight, polymer-surface interaction, and polymer structure (branching),[35–42,43] leading to high experimental cost and time. The application of computer simulations and polymer theory can significantly aid in the rational design of these materials.[24,44,45] Polymer theory approaches, such as self-consistent field theory, have provided predictions for the free energy of clay exfoliation based on a mean-field approximation,[46–51] with chemically specific details of PNC components typically accounted for at the level of a single parameter for each interaction pair.[52] All atom (AA) simulations based on highly accurate pair potentials allow almost exact representation of local interactions and have been applied to study morphology evolution in a diverse set of model PNCs.[53–55,56] However, long relaxation times associated with polymer dynamics and the requirement of large system sizes for highly anisotropic particles limit the utilization of AA simulations for several PNCs of interest.[57–60]

Coarse-grained (CG) simulations can bridge the gap between experimentally relevant and computationally accessible scales for PNCs.[31,61–64] To this end, systematic CG models have been developed for various macromolecules, wherein 4-10 atoms are replaced by a single CG bead with effective pair interaction obtained by combining AA simulation results and experimental data.[65,66] It can involve the use of structure distribution functions as constraints, as done in iterative Boltzmann inversion (IBI) [65,67] and force matching[68] or use of thermodynamic properties such as equilibrium partitioning between polar and apolar solvents as constraints, as done for MARTINI CG force field (CGFF).[66] Chemically specific CG models for PCNCs[29,45,39] can be extremely useful because of the significant effect of the nature of interactions between various components on composite properties and large length- (> 100 nm) and time- (> 1000 ns) scales involved.[69] Suter et al. used IBI-derived CG interaction potential for an MMT-poly(ethylene glycol)/poly(vinyl alcohol) composite to describe the dynamics of polymer intercalation into clay tactoids.[39] However, structure-based CG methods like IBI require extensive optimization across multiple state points and can still exhibit significant parameter-space dependencies.[39,66,67] On the other hand, single interaction parameter (e.g., the Flory χ-parameter) based approaches provide better temperature transferability but fail to capture specific molecular interactions[72] and show limitations in predicting properties of complex systems.[73]

MARTINI CGFF, which is based on defining bead types for individual functional groups making up the polymer or the nanoparticle, can offer good transferability across a wide range of state points with minimal dependence on chemical composition. In MARTINIFF, 3-4 atoms are represented by a single bead, with interaction parameters optimized to accurately reproduce condensed phase densities and partitioning free energies between aqueous and organic phases.[74,66] Initially developed for lipid and biomolecules, it has since been shown to provide excellent representation for several polymer systems.[75–77] Some studies have shown high accuracy of MARTINIFF parameters for a given species across systems involving variations in temperature, solvent, or particle functionalization.[76] For example, the MARTINIFF parameters obtained by reproducing target structural and thermodynamic properties of PE in united atom simulations under various conditions (1,2,4-trichlorobenzene as a good solvent, water as a bad solvent, and melt conditions), showed good agreement with united atom results for both local structure in a PE-polypropylene (PP) blend and free energy profiles of PE dimers penetrating phospholipid membranes from the aqueous phase. The same parameter set also accurately reproduced the structural, thermodynamic, and dynamic behavior of PE chains near the tetramethylammonium (TMA)-MMT surface across multiple temperature conditions, demonstrating robust temperature and chemical space transferability of the model.[77,78] Other studies have also reported the utility of MARTINIFF for modeling polymers near surfaces.[79,80] However, some refinement in self-interaction[81] or cross-interaction[82,83] has been shown to be required for accurately modelling these heterogeneous systems.

In this work, we have extended MARTINIFF to investigate morphology and structure-property relationships in a PE—TMA-MMT composite with polyethylene-polyethylene glycol block copolymer (PE-b-PEG) as a compatibilizer. The CG bead types for edge groups of MMT were parameterized from the orientation-dependent cleavage energy of clay particles, augmenting the previously parameterized CGFF for MMT clays based on a new L-type MARTINI bead.[78] Long CG simulations were used to characterize slow assembly and re-distribution of the compatibilizer. System properties dependent sensitively on atomic packing and interactions, such as the Young's modulus and the glass transition temperature, were determined by backmapping the CG system to AA resolution.

## 2. Methods

TMA-MMT clay particle has a unit cell formula $[TMA]_{0.33}[Si_4O_8][Al_{1.667}Mg_{0.333}O_2(OH)_2]$, involving replacement of MMT inter-gallery ions by the TMA ions. Isomorphic substitutions of $Al^{3+}$ by $Mg^{2+}$ in the octahedral layer were chosen at random, ensuring there are no Mg–(OH)$_2$–Mg bonds, to obtain a charge exchange capacity (CEC) of 91 mmol/100 g.[84,85] Polymer matrix was made of PE (PE80, molecular weight of 2240 $\frac{g}{mol}$), with PE-b-PEG block copolymer ($CH_3(CH_2CH_2)_{16}(OCH_2CH_2)_3OH$, molecular weight: 600 $\frac{g}{mol}$) as a compatibilizer. PE80 is close to the entanglement molecular weight of 2200 $\frac{g}{mol}$ at 300 K (in the infinite chain size limit),[86] thus allowing us to optimize the need for a sufficiently large polymer chain and a reasonable simulation time for sampling equilibrium trajectories.

All MD simulations were carried out using GROMACS 2021.[87,88] Chemdraw[89] was used for generating the initial conformation of polymer chains and Packmol[90] was used for creating the initial box configuration. For AA simulations, CHARMM36 FF[91,92] was used for PE, PE-b-PEG and [TMA]$^+$, and INTERFACE FF[85,93] for the clay particle, with cross-interactions defined using standard mixing rules.[85] For CG simulations, MARTINIFF[66] parameters for PE and PE-b-PEG were taken from Panizon et al.[77] and Rossi et al.,[94] respectively. The MARTINIFF parameters for clay inner beads and clay-polymer cross-interactions were taken from our previous work.[78] The self- and cross-interactions involving the clay edge beads were parameterized in this work, while the clay sheet bonded interactions were re-parameterized to improve agreement with TMA-MMT Young's modulus.

## 2.1. Model development for CG TMA-MMT

Si, Al, or Mg atoms are connected by M–O–M bonds (M = Si, Al, or Mg) in MMT. Accordingly, CG MMT beads were defined as $SiO_2$ or $MO_2H$ (M = Al or Mg), with center at Si, Al, or Mg. This representation preserved the 2:1 layered structure of phyllosilicates through interlayer and intralayer bonds between CG beads. TMA ion [$N^+(CH_3)_4$] was modeled by a regular-sized $Q_0$ bead. MMT beads were modeled using smaller S-type beads, found to be suitable for ring structures in MARTINI parameterization (mapping scheme in Supplementary Information Figure S1).[66,76,95] The CG potential is given as per equation 1,

$$U(r_{ij}) = 4\varepsilon_{ij}\left[\left(\frac{\sigma_{ij}}{r_{ij}}\right)^{12} - \left(\frac{\sigma_{ij}}{r_{ij}}\right)^{6}\right] + \frac{q_i q_j}{4\pi\varepsilon_0 \varepsilon_{rel} r_{ij}} + \frac{1}{2}k_b(b_{ij} - b_0)^2$$

$$+ \frac{1}{2}k_\theta(cos\theta_{ijk} - cos\theta_0)^2 \quad (1)$$

where, $\varepsilon_{ij}$, $\sigma_{ij}$, $q_i$, and $\varepsilon_{rel}$ are interaction strength, bead diameter, bead charge, and relative dielectric constant, respectively. $k_b$ and $k_\theta$ are force constants for bond ($b_{ij}$) and angle ($\theta_{ijk}$) potentials, respectively. $k_b$ and $k_\theta$ of MMT were determined by using Young's modulus and bending stiffness as constraints, obtained from AA simulations of a periodic TMA-MMT sheet (10.35 nm × 10.67 nm) at 300 K. The equilibrium statistics on bond and angle distribution for effective CG beads were obtained from a 10 ns AA simulation of a TMA-MMT sheet.

Within MARTINIFF, the CG beads are categorized as polar (P), non-polar (N), apolar (C), and charged (Q). The LJ well-depth for a given pair of chemical building blocks (beads), $\varepsilon_{ij}$, is determined by calibrating against experimental thermodynamic data, such as oil/water partitioning coefficients. However, the oil/water partitioning coefficients are not available for the clay particle, and role of high covalent coordination in significantly lowering the dispersion interactions[85] precludes use of individual building blocks of clay for determination of partition coefficients. Therefore, we used clay basal spacing and tactoid cleavage energy, experimental measurements are available for both, to parametrize $\varepsilon_{ij}$ for CG clay beads. The cleavage free energy was obtained by considering a slow equilibrium cleavage process allowing for full surface reconstruction of the clay particle.[96] To identify the correct bead type, the cleavage energy was divided into polar (electrostatic) and dispersive (Lifschitz-van der Waals) contributions as per the classical surface theory[97] (methods for determination of Young's modulus, bending stiffness, CG interaction parameters, and calculation of partitioning are available in Supplementary Information Sections S1 and S2).

*TMA-MMT Edge Bead Parameterization*

AA simulation of finite MMT sheets was used to calculate basal spacing and cleavage energies of two sheets in different orientations. A 10.35 nm × 10.67 nm TMA-MMT (12836 atoms in 24 × 24 × 1 array of unit cells) cleaved at the 010 surface was constructed from an equilibrated 2.59 nm × 2.71 nm periodic TMA-MMT sheet. The dangling bonds on each solvent-accessible tetrahedral-octahedral-tetrahedral (TOT) unit at the edge require two water molecules to be saturated, with edge Al, Mg, and Si atoms saturated with $OH^-$ and edge O atoms with $H^+$. TMA ions were moved 1 nm away along the normal to 001 surfaces, five different initial arrangements were created, and MD simulations were continued till the potential energy converged. The lowest potential energy structure was used for subsequent simulations.

This equilibrated AA sheet (Figure 1(a)) was used to obtain an initial configuration for the CG TMA-MMT sheet (Figure 1(b)) with 3040 beads, comprising 2616 inner MMT, 264 edge MMT, and 160 TMA beads. $\varepsilon_{ij}$ for clay inner beads were determined in our earlier work,[78] wherein a new category of L-type beads was defined for the clay beads (e.g. $AlO_2H$, $SiO_2$) with MARTINI levels in the range $SLN_0$-$SLP_1$. The edge beads are expected to have higher polarity because of the presence of either one or two hydroxyl groups, and therefore, test CG simulations were done by assigning all levels in the range $SLP_1$-$SLP_5$. The cleavage energy was calculated for two different arrangements, with either 001-001 (parallel arrangement) or 001-010 (perpendicular arrangement) surfaces in contact (see Supplementary Information Figure S2). The difference in total potential energy between the contact and the separated (two clay particles separated by 10 nm) configurations was used to estimate the cleavage energy. Position restraints in $x$ and $y$ direction were imposed on both clay sheets whereas sheets were allowed to move in the $z$ direction. No position restraints were applied on the TMA ions.

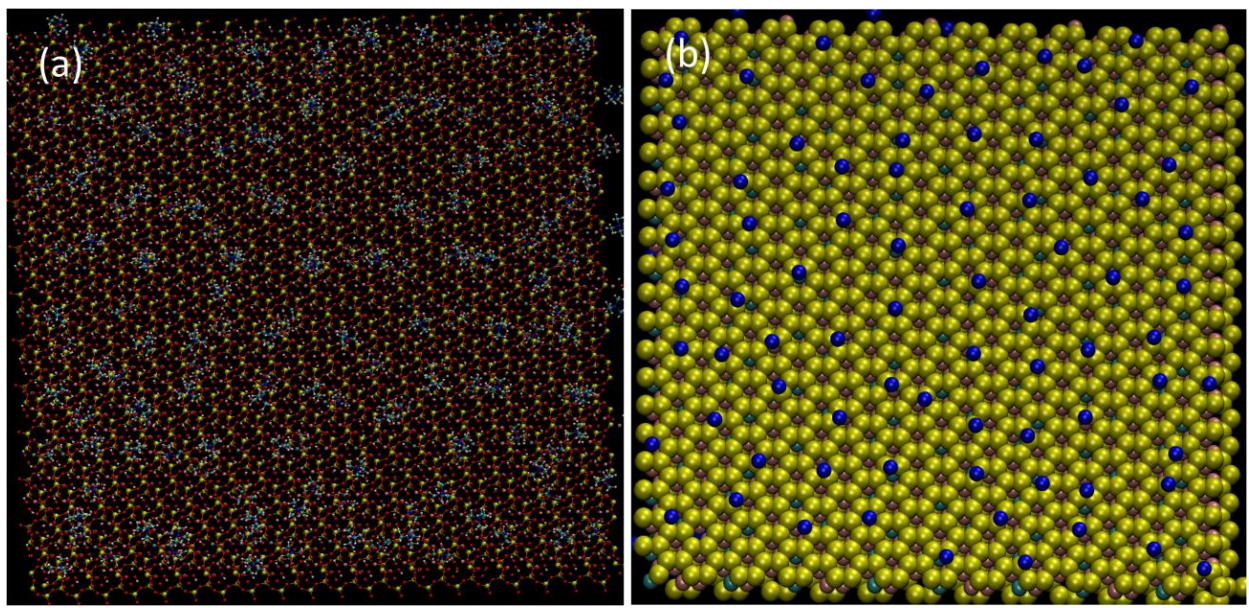

**Figure 1:** AA and CG representation of TMA-MMT sheet (10.35 nm × 10.67 nm) cleaved at the 010 surface. (a) AA representation, with Al/Mg, Si, O, N, C and H shown as pink, yellow, red, blue, cyan and white beads, respectively. (b) CG representation, with $AlO_2H$ or $AlO_2H(OH)$, $MgO_2H$ or $MgO_2H(OH)$, $SiO_2$ or $SiO_2(OH)$ and TMA beads shown as pink, cyan, yellow, and blue, respectively.

Simulations were performed in NVT ensemble at 300 K using 1 fs time step for the AA model and 10 fs for the CG model. Velocity rescale thermostat ($\tau$ = 0.1 ps for AA and 1 ps for CG) was used to maintain the temperature of the system. A 2 ns simulation was carried for both the contact and separated configurations, with last 500 ps used for the determination of average potential energy.

**2.2 Long timescale MD simulation of polymer-clay nanocomposites**

Long timescale and large lengthscale simulations of a representative PCNC were performed to investigate the effect of nanoparticle configuration and composite morphology on PCNC properties.

### 2.2.1. Model development and simulation of PCNCs and melt

The PCNC systems were prepared in Packmol[90] by random insertion of 2000 PE80 and 550 PE-b-PEG molecules around three 10.35 nm × 10.67 nm TMA-MMT sheets (equivalent to 10% w/w concentration in PCNC) placed in two different configurations, one with clay sheets placed in a random orientation in an exfoliated state (referred to as PCNC1) and other with a single clay tactoid placed at the box center (referred to as PCNC2) in a 30 nm × 30 nm × 30 nm box. Further, a melt system comprising only PE80 and PE-b-PEG was also prepared. Details of polymer-clay systems simulated here are given in Table 1.

**Table 1. Details of simulation systems.** PCNC1 refers to a system where clay sheets are placed in an exfoliated state within the polymer melt, while PCNC2 represents a system in which a clay tactoid is placed within the polymer melt. The specific properties calculated from each simulation are specified in the last column: cleavage energy ($\sigma$), basal spacing ($d$), coordination number (CN), radial distribution function ($g(r)$), radius of gyration ($R_g$), end-to-end distance ($R_e$), conformational entropy ($s_c$), diffusion coefficient ($D$), Young's modulus ($E$), and glass transition temperature ($T_g$), respectively.

| System | Components (number) | T (K) | Resolution | Simulation time (ns) | Calculated Property |
|---|---|---|---|---|---|
| TMA-MMT | TMA-MMT (2) | 300 | AA, CG | 2-60 | $\sigma, d, E$ |
| PE-TMA-MMT | PE80 (720), TMA-MMT (1) | 450 | AA, CG | 100 | $R_g, R_e, s_c, D$ |
| Melt | PE80 (610), PE16-b-PEG3 (2210) | 450 | CG | 5000 | CN, $g(r)$, $R_g, R_e, s_c$ |
| PCNC1, PCNC2 | PE80 (550), PE16-b-PEG3 (2000), TMA-MMT (3) | 450 | CG | 600 | CN, $g(r)$, $R_g, R_e, s_c$ |
| Melt1, Melt2 | PE80 (610), PE16-b-PEG3 (2210) | 140 | AA | 60-120 | $E, T_g$ |
| PCNC1, PCNC2 | PE80 (550), PE16-b-PEG3 (2000), TMA-MMT (3) | 140 | AA | 60-120 | $E, T_g$ |

The system density (~ 0.3-0.4 of target density obtained using Packmol) was increased in several isotropic compressions (5%) – energy minimization (steepest descent EM run till the maximum force became less than $1000 \frac{kJ}{(mol)(nm)}$) cycles to ~ $770 \frac{kg}{m^3}$ (for comparison, PE80 melt has a density of $782 \frac{kg}{m^3}$ at 450 K[77,78]), corresponding to a box size of 22.6 nm × 22.6 nm × 22.6 nm. These systems were heated from 0 K to 450 K in 400 ps by coupling to a V-rescale thermostat ($\tau$ = 0.1 ps). An 80 ns simulation with isotropic pressure coupling at 1 bar and periodic temperature annealing (each cycle 4 ns) between 450 K and 550 K was used to obtain a box of 22.09 nm × 22.09 nm × 22.09 nm in case of PCNC1 (for additional details see Supplementary Information Section S3). In the last 10 cycles, center-of-mass (COM) MSD reached up to $(3.5R_g)^2$, and the polymer density and $R_g$ changed by less than 2%. Thus obtained configurations of PCNCs and melt systems were run at 1 bar and 450 K for 600 ns and 5000 ns, respectively, for property calculation. The methodology for computing structural, thermodynamic, and dynamical properties is outlined in Supplementary Information Section S3.1.

*Calculation of preferential interaction coefficients ($\Gamma_{XP}$)*

The preferential interaction coefficient ($\Gamma_{XP}$) of species X was calculated as per equation (2).[98–100]

$$\Gamma_{XP}(R) = \langle n_{XP}(r<R) - \frac{n_X - n_{XP}(r<R)}{n_P - n_{PP}(r<R)} n_{PP}(r<R) \rangle_{\tau_{run}} \quad (2)$$

Here, $\langle\ \rangle_{\tau_{run}}$ is the time average over the equilibrium portion of length $\tau_{run}$ of MD trajectory, $n_X$ and $n_P$ are the total number of co-solvent and solvent molecules in the simulation box. $n_{XP}$ ($r < R$) and $n_{PP}$ ($r < R$) are the number of co-solvent and solvent molecules, respectively, for which the monomer center of mass or CG bead center falls within a distance R from the van der Waals surface of the closest MMT bead (determined using the *gmx select* tool of Gromacs 2021[87,88]). Here, monomers/beads of PE polymer chains were considered as solvent molecules, and either the $PE_b$ or $PEG_b$ monomers/beads of PE-b-PEG were considered as co-solvent molecules.

*Coordination numbers*

Coordination numbers (CN) of all possible pairs among clay beads, PE, $PE_b$, and $PEG_b$ were calculated from the radial distribution functions (RDF, $g_{\alpha,\beta}(r)$) of each pair (Equation (3)).

$$CN = \rho_\beta \int_0^{r_{\text{cut}}} 2\pi r^2 g_{\alpha,\beta}(r) dr \qquad (3)$$

Here, $\rho_\beta$ is the number density of $\beta$ species and $r_{\text{cut}}$ is cut-off distance for CN calculations, taken as the largest value for the second minima in $g_{\alpha,\beta}(r)$ (~ 1.2 nm, see Supplementary Information Figures S6, S7, and S8 for the pair RDFs). The contribution to $g_{\alpha,\beta}(r)$ from the neighboring bonded atoms were removed using three and nine exclusions for the polymer and MMT sheet, respectively. For all three systems, the RDFs were computed as the average over all 10 ns blocks taken at 50 ns intervals. For the composite system, CN converged after 300 ns (Figure 4), and therefore, averages were calculated from the last 200 ns trajectory. In the melt system, PE-b-PEG undergoes gradual clustering, with RDFs showing a convergence after 4000 ns (see Supplementary Information Figure S8). The averages for the melt were calculated from the final 1000 ns trajectory.

*Cluster size and distribution of PEG beads near the clay surface*

Cluster analysis was performed using the *gmx clustsize* tool of Gromacs 2021[88] to evaluate the time evolution of the average cluster size and the number of $PEG_b$ bead clusters in both PCNCs over the 600 ns production trajectories. A bead was considered part of a cluster if it was within 0.45 nm of any other bead in the cluster. The 600 ns trajectory was divided into fifteen 40 ns blocks, and cluster analysis was performed separately on each block to characterize time evolution of clusters.

**2.3. Backmapping of melt and PCNCs from CG to AA structure**

For PCNCs, CG TMA-MMT was mapped to the atomistic model by placing Si, Al, or Mg at the center of the corresponding CG bead (e.g, Si placed at $SiO_2$ or $SiO_2(OH)$), followed by stepwise addition of O and H atoms. Figure 2 shows a schematic of the 3-step backmapping procedure for TMA-MMT, detailed as below.

*Step 1: reference sheet alignment*: Each CG TMA-MMT sheet was mapped separately to AA, using an equilibrated AA TMA-MMT as reference (referred to as, rAA(TMA-MMT)). Rotational and translational fit of rAA(TMA-MMT) to the target CG sheet (using the positions of the Si, Al, Mg atoms) was done to ensure that two tetrahedral layers of rAA(TMA-MMT) are correctly aligned with corresponding tetrahedral layers of the target CG sheet.

*Step 2: sheet "backbone" frame construction*: Al, Mg, and Si atoms (M) were placed at the center of corresponding CG beads, representing a direct inversion of the coarse-graining step.

*Step 3: addition of oxygen atoms and hydroxyl group:* For each M atom, two closest oxygen atoms in rAA(TMA-MMT) (say, $O_1$ and $O_2$) were identified. The $MO_1$, $MO_2$, and OH (for hydroxyl group) vectors were determined from rAA(TMA-MMT). Then, $O_1$, $O_2$, and H atoms were introduced into the target CG sheet by adding these vectors to the corresponding M atom. This step generated a backmapped AA coordinate file of the target CG sheet. Thus obtained TMA-MMT structure was relaxed in a single energy minimization step.

Separately, PE and PE-b-PEG polymers were converted to an AA representation using the procedure described by Wassenaar et. al.[101] (more details in Supplementary Information Section

S4). The definitions of the weighted geometrical centers and refining rules used for polymer backmapping are provided in the form of readable mapping files that can be directly used with the *backward.py* script available on the CG Martini webpage.

The individual TMA-MMT sheets and the full polymer component coordinate files were merged to obtain the AA resolution PCNC structure. The melt CG structure was converted to AA structure using the polymer backmapping protocol.

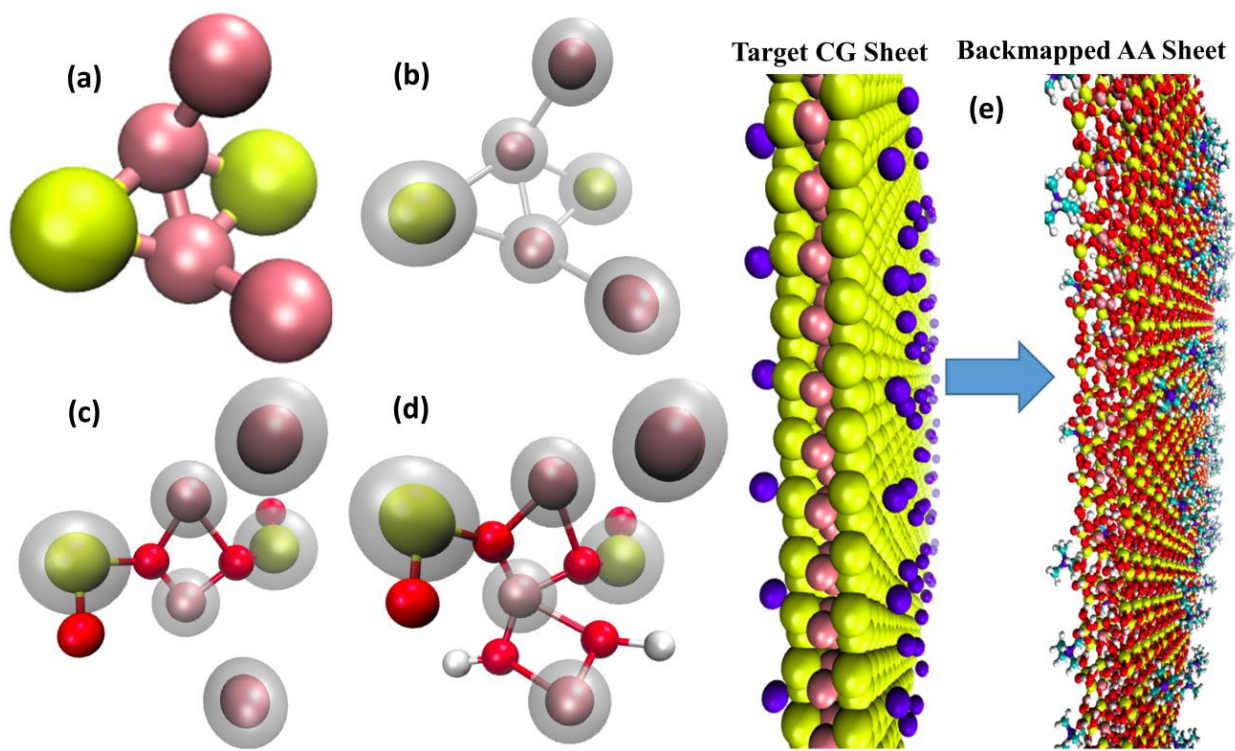

**Figure 2**: CG to AA backmapping of TMA-MMT sheet. (a) CG representation, (b) placement of heavy atoms (M = Al or Mg and Si) at the center of mass of CG beads, (c) and (d) insertion of O and H atoms by adding MO vectors to the corresponding Al, Mg, and Si atoms and OH vectors to the O atom in the target CG sheet, respectively. (e) AA sheet obtained from CG sheet after backmapping and energy minimization. Color scheme: $SLN_0$ (pink), $SLN_a$ (yellow), $SLP_2$ (yellow), Al (pink), Si (yellow), O (red), TMA (blue). The grey circles in (b), (c), and (d) represent the cross-section of the original CG bead.

### 2.3.1 AA simulation of backmapped melt and PCNCs

Five equally spaced (50 ns interval) structures from last 250 ns CG trajectories were backmapped to AA resolution. Each structure was relaxed using steepest-descent energy minimization, heated to 450 K in a 250 ps NVT simulation, and was further relaxed in a 4 ns NPT simulation at 450 K and 1 bar (Berendsen barostat with $\tau_P = 0.5$ ps for first 2 ns and Parrinello-Rahman barostat with $\tau_P = 3$ ps for next 2 ns). Figure S15 shows that backmapped structures (5 each for four distinct morphologies) reach a potential energy plateau in given time. The CG and the corresponding backmapped AA structures are shown in Figure S16. The structures obtained at the end of 4 ns simulation were isobarically cooled to 140 K with a cooling rate of $5\frac{K}{ns}$ at 1 bar. The cooling rate was chosen based on the accurate reproduction of the glass transition temperature and the Young's modulus in smaller system consisting of thirty PE chains (details in Supplementary Information Section S5).

*Property estimation*

The temperature-density relation obtained from the isobaric cooling simulation was used to determine the glass transition temperature ($T_g$) and this estimate was extrapolated to the experimental cooling rates as per the Williams−Landel−Ferry (WLF) equation.[102] The structures at 140 K and 200 K were subject to a series of uniaxial tensile simulations to determine the Young's modulus ($E_i$). Additional details are given in Supplementary Information Section S5.1.

**3. Results and Discussion**

**3.1 Parameterization and validation of CG model for clay edge beads**

The CG parameters for saturated TMA-MMT edge beads were obtained by using select properties from the reference AA simulation as constraints. These were combined with available CG parameters for clay inner beads and PE polymer to validate the CG parameter set by comparing the structural, thermodynamic, and dynamical properties obtained from atomistic simulations.

*CG bonded potential for TMA-MMT*

Previously[78] parameterized bonded potential for CG TMA-MMT was based on the elastic and bending moduli of the corresponding AA sheet obtained using the INTERFACE FF.[85] This parameter set led to small buckling in CG simulations of finite sheets (Supplementary Information Figure S3). Further, Zartman et al.[103] recommended decreasing the AA bonded potential stiffness in INTERFACE FF by 40% to improve agreement with the MMT elastic modulus of 160 GPa obtained using density functional theory calculations. We, therefore, decreased the CG bonded potential stiffness ($k_b$ and $k_\theta$ for MMT beads in equation (1)) by 40% and obtained elastic moduli of $E_x = 170.5$ GPa and $E_y = 170.8$ GPa, in excellent agreement with corresponding AA simulation (revised potential[103]) values of 171.9 GPa and 172.4 GPa, respectively. The bending stiffness of $\frac{D_x}{h^2} = 15.43$ N m$^{-1}$ and $\frac{D_y}{h^2} = 15.47$ N m$^{-1}$ was also in excellent agreement with the corresponding AA estimates of 15.32 N m$^{-1}$ and 15.36 N m$^{-1}$, respectively.

**3.1.1** *CG Non-bonded potential for TMA-MMT edge beads*

The average basal spacing ($d$) and cleavage energy ($\sigma$) of TMA-MMT sheets in parallel and perpendicular arrangements (see Supplementary Information Figure S2) were calculated from a 2 ns AA simulation. We obtained $\sigma = -107.8 \, \frac{\text{mJ}}{\text{m}^2}$ for finite sheets in a parallel arrangement, with edge polar groups making this value slightly more negative than $\sigma = -105.6 \, \frac{\text{mJ}}{\text{m}^2}$ obtained for the periodic TMA-MMT.[78] The polar (electrostatic) ($\sigma_p$) and dispersive (Lifschitz-van der Waals) ($\sigma_d$) contributions to the cleavage energy were calculated to optimize the edge bead type assignments (sample calculation in Supplementary Information Section S1).

**Table 2.** MARTINI bead types used for chemical building blocks of finite TMA-MMT

| Chemical group | MARTINI type | Polarity |
|---|---|---|
| $SiO_2$ | [†]$SLN_a$, [††]$SN_a$ | Non-polar h-bond acceptor |
| $AlO_2H$ | [†]$SLN_0$, [††]$SN_0$ | Non-polar |
| $MgO_2H$ | [†]$SLQ_0$, [††]$SQ_0$ | Charged |
| $SiO_2(OH)$ | [†]$SLP_2$, [††]$SP_2$ | Polar |
| $Si(OH)_2$ | [†]$SLP_2$, [††]$SP_2$ | Polar |
| $AlO_2H(OH)$ | [†]$SLP_2$, [††]$SP_2$ | Polar |
| $MgO_2H(OH)$ | [†]$SLQ_0$, [††]$SQ_0$ | Charged |
| TMA | [†]$LQ_0$, [††]$Q_0$ | Charged |

[†]: Martini type for clay-clay interaction
[††]: Martini type for clay-polymer/ion interaction

Previously,[78] the surface tetrahedral bead [$SiO_2$] was modeled as $SLN_a$ (non-polar h-bond acceptor), as isomorphic substitutions of $Al^{3+} \rightarrow Mg^{2+}$ within the octahedral layer have been shown to enhance the polarity of surface [$SiO_2$] blocks.[104] The octahedral layer beads, [$AlO_2H$] and [$MgO_2H$]$^-$ were modeled as $SLN_0$ (non-polar) and $SLQ_0$ (charged), respectively. This assignment provided an excellent agreement between AA and CG models for polar and dispersive contributions to the cleavage energy of TMA-MMT. Here, using $SLN_a$ type for all edge beads gave a perpendicular arrangement $\sigma$ of $-137.8 \frac{mJ}{m^2}$, a significant underestimate for the corresponding AA value of $-157.6 \frac{mJ}{m^2}$. To account for the higher polarity of edge beads consisting of 1-2 hydroxyl groups, CG simulations were rerun using higher polarity beads in the range $SLP_1$ to SLP$_5$. The uncharged edge beads, viz. ([AlO$_2$H(OH)], [SiO$_2$(OH)], and [Si(OH)$_2$]), have similar functional groups (both bridging and hydroxyl groups are present) and similar local connectivity, and therefore, were assigned the same S-type bead to limit the parameter space. The edge [MgO$_2$H(OH)] bead carries a -1 charge and was modeled as SLQ$_0$ (final bead type assignments given in Table 2). Table 3 shows that the SLP$_2$ type for edge beads provided the best concurrent estimates for $\sigma$, $\sigma_d$, $\sigma_p$, and $d$. The difference in the cleavage energy of AA and CG sheets was only 5% (parallel) and 2% (perpendicular) for the dispersive component and 8% (both parallel and perpendicular) for the polar component. The basal spacings obtained from CG essentially exactly match with corresponding AA values. This optimal P$_2$ type for edge beads lies between the alcohol group (P1) and the diol groups (P$_3$ or P$_4$), and thus, corresponds well to the polarity of functional groups.[66]

**Table 3:** Cleavage energy and basal spacing of TMA-MMT clay sheet. $\sigma, \sigma_d, \sigma_p$, and $d$ were calculated for finite clay sheets in parallel and perpendicular arrangements, reported here as averages from 2 ns MD simulations at 300 K and 1 bar.

| Properties | AA | CG |
|---|---|---|
| Clay sheets in parallel arrangement | | |
| $\sigma \left(\frac{mJ}{m^2}\right)$ | -107.8 | -116.2 |
| $\sigma_d, \sigma_d \left(\frac{mJ}{m^2}\right)$ | -20.47, -87.31 | -21.5, -94.70 |
| d (nm) | 1.30 | 1.299 |
| Clay sheets in perpendicular arrangement | | |
| $\sigma \left(\frac{mJ}{m^2}\right)$ | -157.6 | -163.1 |
| $\sigma_d, \sigma_d \left(\frac{mJ}{m^2}\right)$ | -44.37, -117 | -45.3, -124.4 |
| d (nm) | 0.51 | 0.51 |

**3.1.2 Structural, thermodynamic, and dynamical properties of clay-polymer systems.**

A model polymer-clay nanocomposite consisting of 720 PE80 chains around a single TMA-MMT (10.35 nm × 10.67 nm) (see Supplementary Information Figure S5) was used for an extensive comparison of properties determined from the CG and the AA simulations. The cross-interaction parameters for clay and polymer bead interactions were taken as per MARTINI v2 assignments ($\varepsilon_{ij}$ given in Supplementary Information Table S3). There was an excellent agreement between AA and CG models (Table 4): radius of gyration $R_g$ within 3%, end to end distance $R_e$ within 4%, conformational entropy per bead $s_c$ within 3%, and normalized diffusion coefficient $\frac{D}{D_0}$ within 4%. In both AA and CG models, $R_g$ and $R_e$ show a small increase of ~1% and $s_c$ a small decrease of ~2% for this PCNC system compared to the value in the melt, indicating only a small effect of the clay surface on the PE chains, which possibly acts as a neutral (no preferential attraction or repulsion) substrate for PE and is in agreement with other experimental and simulation studies for polymers near surfaces.[105–107] The self-diffusion coefficient of the polymer Kuhn segment in AA and CG simulation was obtained to be $6.09 \times 10^{-7}$ and $8.14 \times 10^{-7}$ $\frac{cm^2}{s}$ at 450 K, respectively. The former compares fairly well with chain diffusivity of $5.4 \times 10^{-7} \frac{cm^2}{s}$ for a PE88 melt at 454 K obtained using a modified nuclear magnetic resonance (NMR) field gradient technique.[108] A ~ 30% higher diffusion coefficient in CG PCNC (w.r.t AA) is in line with other studies[76,78] and results from reduced atomic friction in the CG system. We, therefore, used the normalized diffusion coefficient $\left(\frac{D}{D_0}\right)$ to evaluate dynamic properties predictions in CG system. $\frac{D}{D_0}$ was obtained to be 0.84 and 0.87 in AA and CG PCNC, respectively, again representative of a weak interaction between PE and TMA-MMT, in agreement with simulation and experimental studies of polymers near solid surface.[109,110]

Table 4 Comparison of structural, thermodynamic, and dynamical properties for AA and CG PE—TMA-MMT system having single clay sheet. Radius of gyration ($R_g$) and end-to-end distance ($R_e$) of polymer chains, the normalized diffusion coefficient of the Kuhn segments $\left(\frac{D}{D_0}\right)$, and conformational entropy per CG bead ($s_c$) were calculated at 450 K and 1 bar. Average and standard deviation was calculated over 100 ns simulation using 20 ns blocks.

|  | AA | CG |
|---|---|---|
| $\langle R_g \rangle$ (nm) | 2.17 ± 0.15 | 2.24 ± 0.18 |
| $\langle R_e \rangle$ (nm) | 5.11 ± 0.23 | 4.88 ± 0.25 |
| $\frac{D}{D_0}$ | 0.87 | 0.84 |
| $s_c \left(\frac{J}{mol.K}\right)$ | 58.18 | 56.92 |

## 3.2 Largescale simulation of melt and PCNCs

The developed CG parameters were used to investigate slow morphology evolution in the melt (PE80 and PE-b-PEG blend) and the two composite (PCNC1 & PCNC2, as defined in Table 1) systems and to obtain near-equilibrium morphologies for further property determination at AA resolution (Figure 3 shows the overall workflow employed in this study).

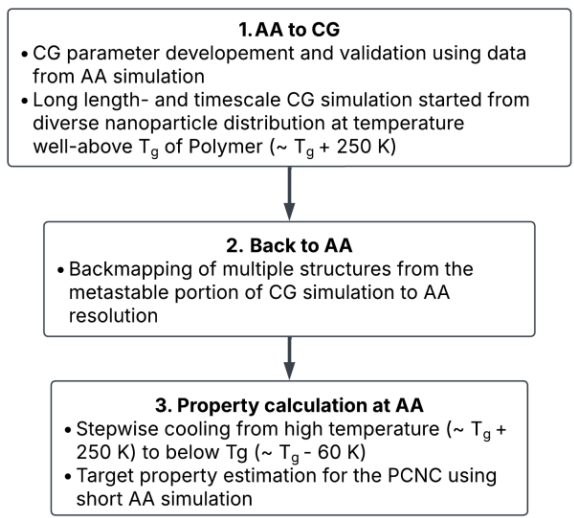

**Figure 3**: The multiscale workflow used for melt and PCNC systems.

### 3.2.1 Nanofiller dispersion and microstructure evolution

The microstructures evolution in different systems (PCNC1, PCNC2 and melt) was investigated using pair coordination numbers (CN) (see Equation 3). Figures 4 and 5 show that all CNs converge in 4000 ns for the melt and 300 ns for PCNCs, respectively, with the remaining trajectory used to calculate average CNs and RDFs (see Supplementary Information Table S4 for trajectory length used and Figures S6, S7, and S8 for RDFs). Figure 4(a) shows that due to the large fraction of PE, changes in CN relative to PE remain small. However, a slight increase in PE self-interaction is observed, with PE-PE CN increasing from 53.63 to 54.67 and PE-PEG$_b$ and PE-PE$_b$ decreasing from 0.801 to 0.272 and 2.55 to 1.68, respectively. This effect is more pronounced in the CN of PEG$_b$ (Figure 4b) and PE$_b$ (Figure 4c), where the decline in cross-interaction with PE is more clearly visible. PE$_b$-PE CN significantly greater than PE$_b$-PE$_b$ CN (Figure 4(c)) at small times implies that the initial structure has a high dispersion of PE-b-PEG block copolymer in PE matrix. Over the first 4 $\mu s$ trajectory, the PEG$_b$-PEG$_b$ CN increased from 29.55 to 50.58 and PE$_b$-PE$_b$ CN increased from 7.33 to 18.76, implying the formation of micelle-like block copolymer clusters (for example, as seen in Figure 6(f)). A similar phase separation was evident through the Han curves (between storage and loss modulus) of a blend of ultra-high molecular weight PE and PEG.[111] Another study, employing optical microscopy, reported that PEG was positioned at the interface of bitumen and polyethylene, contributing to a compatibilization effect.[112]

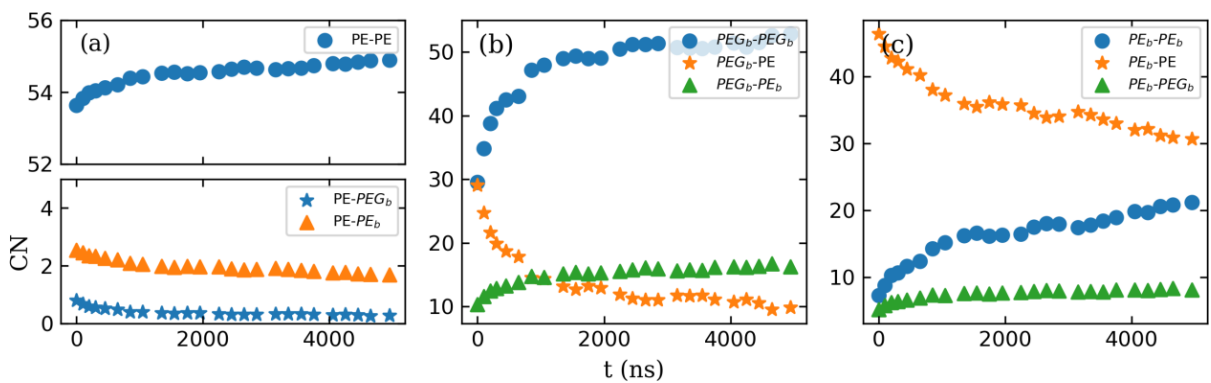

Figure 4: Morphology evolution in the PE/PE-b-PEG melt. Time evolution of the coordination number (CN, calculated up to the second shell) for all CG beads around (a) PE, (b) $PEG_b$, and (c) $PE_b$ in PE/PE-b-PEG melt. The simulation was performed at 450 K and 1 bar.

In this study, we have investigated the role of PE-b-PEG in aiding the dispersion of TMA-MMT in a PE matrix. This provides an alternative to the use of oMMT with large alkyl chain substitutions for achieving a good dispersion of the clay particle, which was found to be important to obtain large improvement in tensile strength, low fractional free volume, and other target properties in these composite systems.[113,114]

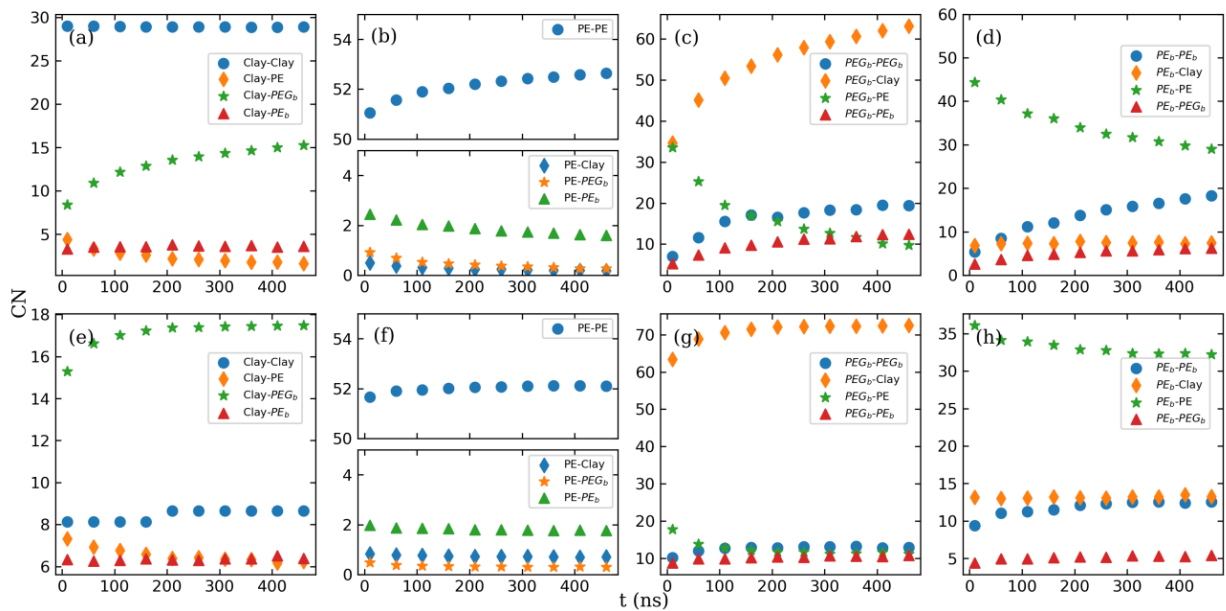

Figure 5: Morphology evolution in PCNCs. (a), (b), (c), and (d) represent time evolution of second shell CN in clay tactoid PCNC (PCNC2) for all possible pairs (clay, PE, $PEG_b$ and $PE_b$) around clay, PE, $PEG_b$, and $PE_b$, respectively. (e), (f), (g) and (h) represent the same quantities in exfoliated clay PCNC (PCNC1).

Figures 5(a), (e) show that $PEG_b$ CN around clay increased from 8.36 to 15.23 in the clay tactoid PCNC (PCNC2) and from 15.29 to 17.48 in the exfoliated clay PCNC (PCNC1), while that for PE and $PE_b$ stayed much smaller, indicating a preferential $PEG_b$-clay interaction. This significant difference can be attributed to the re-orientation of PE-b-PEG chains within the simulation time such that the clay surface is almost fully saturated by $PEG_b$ (see Figures 6(a-d)). This reorientation also lead to a small decrease in $PE_b$-PE CN (Figures 5(d), (h)) and a significant decrease in $PEG_b$-PE CN (Figures 5(c), (g)) in both PCNCs. A similar preferential interaction of PEG with a silicate surface was shown to lead to a decrease in the polymer blend contact angle (from 96° to 65°) on the addition of PE-b-PEG of different lengths (350-mer to 2000-mer) to a PE melt.[115] This preferential migration of PEG to clay surface is primarily driven by the favorable interaction of PEG hydroxyl group with either the bridging oxygen atoms (hydrogen bond acceptors[116]) or the edge hydroxyl groups (hydrogen bond acceptor and donor) of the siloxane surface. These favorable clay-PEG interactions decreased $PEG_b$-$PEG_b$ CN from melt value of 52.53 to 18.92 in PCNC2 and to 13.04 in PCNC1 (see Supplementary Information Table S4). This marked reduction demonstrates effective disruption of $PEG_b$ clusters by the introduction of clay assemblies (see Figure 6 (b) and 6 (d)), and this effect is more prominent in PCNC1 due to the large accessible surface area of clay. The time evolution of $PE_b$-$PE_b$ showed a continuous increase in PCNC2 (Figure 5 (d)), with the average calculated over the final 200 ns trajectory approaching a value close to the melt (see Supplementary Information Table S4). However, the $PE_b$-$PE_b$ CN approached a plateau in the simulation timescale for PCNC1 (Figure 5 (h)), with the average value ~ 60% of that in the melt, again a consequence of effective disruption of block copolymer assembly in the exfoliated PCNC1 system. The clay-PE (Figures 5(a),(e)) and $PE_b$-PE (Figures 5(d),(h)) CNs also approach a plateau, with both values somewhat smaller in PCNC2 compared to those in PCNC1. This larger contact between the matrix PE and compatibilizer-nanoparticle complex implies more effective compatibilization in the PCNC1 system.

A small abrupt increase in clay-clay CN in the exfoliated system ~ 200 ns (Figure 5(e)), also observable from the emergence of a two clay assembly intercalated by PE-PEG copolymer (Figure 6(b)), is indicative of a slow structural evolution in these PCNC systems at scales prohibitively large for an all atom MD simulation. A high self-affinity of oMMT was also evident from the formation of intercalated and aggregated MMT assemblies at high MMT loading in a PE-MMT system.[114] At the same time, an excess of low molecular weight block copolymer species, for example, as seen in the PCNC2 (Figure 6(d)), would lead to a decrease in matrix stiffness (made up of high molecular weight PE) and negatively affect other target properties. Therefore, the MMT loading and PE-b-PEG block copolymer concentration should be optimized depending on the level of clay exfoliation to inhibit clay assembly and to limit the concentration of free block copolymer.

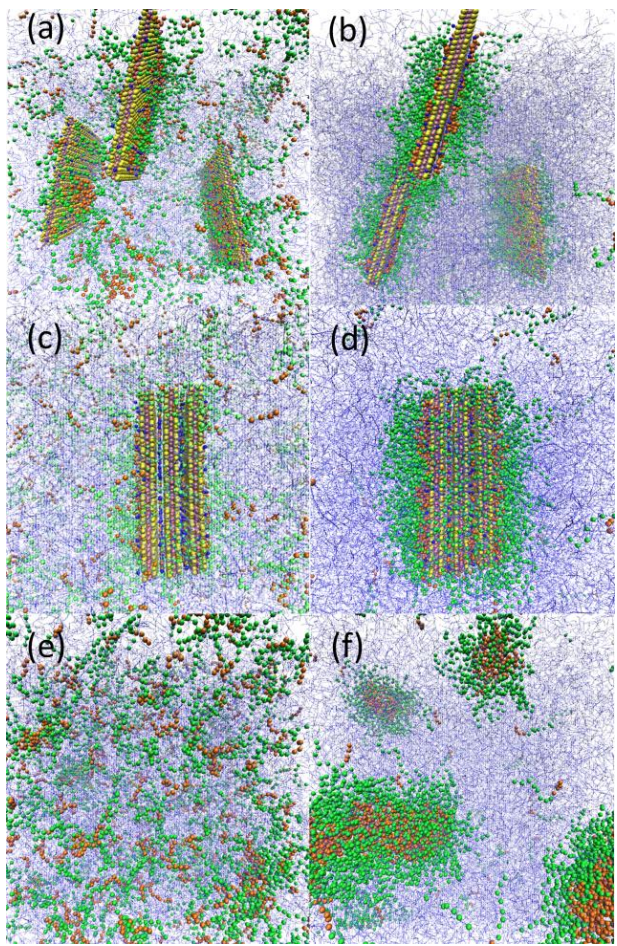

**Figure 6**: Morphology evolution in PCNCs and melt. (a) and (c) Starting structure for the production run (obtained after local equilibration steps) for PCNC1 and PCNC2, respectively. (b) and (d) Final structure obtained at the end of 500 ns simulation for PCNC1 and PCNC2, respectively. (e) and (f) Melt system at 100 ns and 5000 ns, respectively. The yellow, pink and blue spheres denote the CG beads of TMA-MMT as defined in the Figure 1, orange and green spheres denote the PEG and PE beads of PE-b-PEG polymer, and for clarity the PE80 polymer chains are represented by blue lines.

### 3.2.2 Effect of clay dispersion on compatibilizer distribution and assembly

Individual distribution of $PEG_b$ and $PE_b$ components of PE-b-PEG was characterized using preferential interaction coefficients $(\Gamma_{xp})$ computed w.r.t. to the clay surface (shown in Figure 7). A significantly higher $\Gamma_{xp}$ for $PEG_b$ implies preferential solvation of the clay surface by $PEG_b$. The distance between the sheet and $PEG_b$ beads was less than 0.65 nm in PCNC1, indicating essentially a monolayer coverage by $PEG_b$ (also see Figure 6(b)). However, in PCNC2, a lesser clay surface was accessible because of tactoid formation, leading to the formation of a double layer of $PEG_b$. Essentially no $PE_b$ was present in the first layer in both PCNCs, an observation in agreement with self-consistent field theory (SCFT) calculations showing anchoring of the polar part of short-chain surfactants at the clay surface.[47,49]

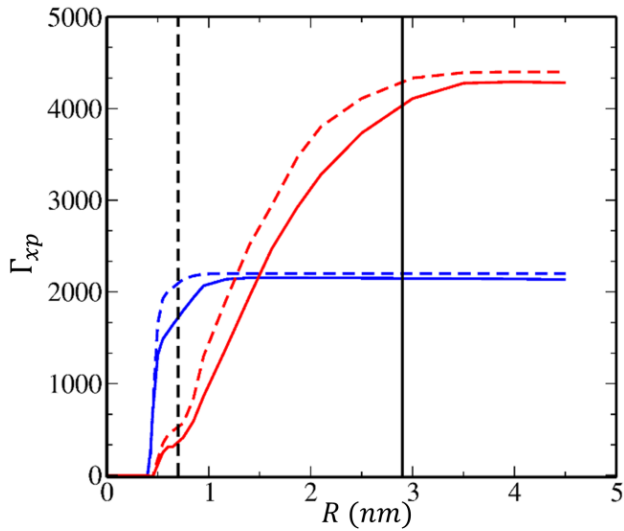

**Figure 7:** Preferential interaction coefficients, Γxp, for $PE_b$ (red) and $PEG_b$ (blue) beads of PE-b-PEG plotted as a function of smallest distance to the sheet, $R$ (see Equation 2). The dashed lines are for the exfoliated state PCNC1 and the solid lines for the clay tactoid PCNC2. The dashed black line is placed at 0.65 nm from the clay surface and the solid black line at $(0.65 + 2R_g)$, where $R_g$ is the radius of gyration of PE beads in PE-b-PEG.

Figure 8(a, b) show that PEG$_b$ association evolved from many small clusters at short times to a smaller number of large clusters, plateauing at an average cluster size (S) of ~ 50 and 45 beads for exfoliated and tactoid PCNCs, respectively. Figure 8(c, d) show that in the initial structure, ~ 18% and 38% PEG$_b$ beads are present as isolated chains (4 bead cluster) and a plateau emerges in the distributions after 400 ns, with ~ 65% and 74% PEG$_b$ beads part of large clusters of size 224-460 for PCNC1 and PCNC2, respectively. Further, there is an almost complete absence of isolated chains in PCNC1. Together, the data from Figures 7 and 8 clearly establish slow, clay induced redistribution of the block copolymer in PCNC systems, with small but important dependence on the nanofiller distribution.

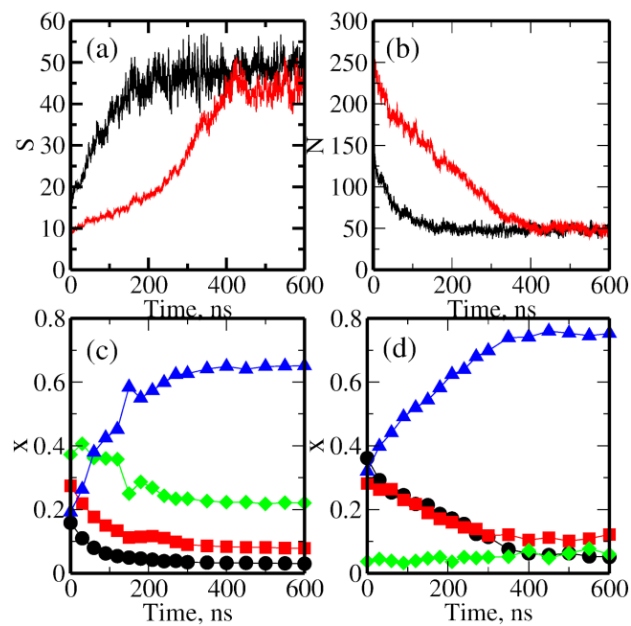

**Figure 8:** PEG clustering at clay surface. Time evolution of (a) average cluster size (S) and (b) number of clusters (N) of PEG beads of PE-b-PEG in CG MD simulation of PCNC1 (black) and PCNC2 (red). Time evolution of the fraction of beads ($X$) in different-sized clusters in (c) PCNC1 and (d) PCNC2. Cluster sizes are grouped as, isolated chains (4 beads) (black circle), small (8-40 beads) (red square), intermediate (44-220 beads) (green diamond), and large (224-460 beads) (blue triangle).

### 3.2.3 Structural, thermodynamical, and dynamic properties of PCNCs

The time evolution of $R_g$, $s_c$, (see Supplementary Information Figure S11 and S12) and CN (Figures 4 and 5) show convergence in the first 300 ns for PCNCs and 4000 ns for melt. PE-b-PEG $s_c$ is ~ 60% of that for PE, resulting primarily from its self-association in both the melt and the composite systems (Table 5) and to a lesser extent from its smaller size (12 monomers versus 40 monomers in PE). For both PE and PE-b-PEG, there is only a marginal change in $R_g$ and $s_c$ in going from the melt to the composite systems. It was noted earlier that the TMA-MMT acts as a weakly interacting surface for PE and the presence of PE-b-PEG at the clay surface in PCNC1,2 further accentuates the bulk-like environment for near-surface PE (PE-PE$_b$ CN higher than PE-clay CN, as seen in Figure 5(b), (f)). For PE-b-PEG, the PEG$_b$ beads are in a similar cluster state in the melt and composites, while the PE$_b$ beads are primarily in coordination with PE$_b$ and PE beads, thus, also preserving the melt-like morphology in the composite system.

A larger effect of the nanofiller was observed on the polymer segment self-diffusion coefficient. The PE diffusion coefficient decreased by a factor of ~0.88 from melt to PCNC2 and to ~0.82 in PCNC1, which can be attributed to the confinement and steric effect of TMA-MMT on PE. Other studies have reported similar observations, where the $R_g$ and $R_e$ of polymer chains was affected only close to the surface or in the first adsorbed layer[107,117] while dynamics were found to be slower for chains at much larger distance from the surface.[105] More recent studies[21,31,109,110] of model polymer–solid systems using an atomistic surface showed that the presence of the solid surface led to significantly slower dynamics. This is in part a direct consequence of the steric effect of a relatively stationary particle reflecting the polymer chain. Overall, the influence of the nanofiller on polymer in the present study is a result of confinement and steric effect of the clay sheets on the PE chains, implying a dominant role of entropic effects in this system with weak nanofiller-polymer interactions.

**Table 5**. Polymer properties in melt and PCNCs. Chain radius of gyration ($R_g$), chain end-to-end distance ($R_e$), and conformational entropy per CG bead ($s_c$) for the two PCNCs (average from last 200 ns) and melt (average from last 1000 ns) calculated at 450 K.

| Polymer Species | Property | CG PCNC1 | CG PCNC2 | CG Melt |
|---|---|---|---|---|
| PE | $\langle R_g \rangle$ (nm) | 2.22 ± 0.005 | 2.19 ± 0.011 | 2.20 ± 0.001 |
|  | $\langle R_e \rangle$ (nm) | 4.98 ± 0.019 | 4.97 ± 0.026 | 4.98 ± 0.023 |
|  | $s_c \left(\frac{J}{mol.K}\right)$ | 49.12±0.008 | 50.64±0.012 | 50.71 ± 0.021 |
|  | $\frac{D}{D_0}$ | 0.82±0.012 | 0.88±0.082 | 1 |
| PE-b-PEG | $\langle R_g \rangle$ (nm) | 1.068±0.001 | 1.079±0.002 | 1.096 ± 0.001 |
|  | $s_c \left(\frac{kJ}{mol}\right)$ | 30.21±0.013 | 30.28±0.024 | 30.82±0.009 |

**3.3 CG Structure Backmapping and Property Calculation**

The CG structures were backmapped to AA resolution for accurate determination of experimentally available properties such as the Young's modulus and the glass transition temperatures. Two distinct CG melt morphologies were taken, viz. melt1 (Figure 6(e)) having a highly dispersed PE-b-PEG in PE obtained after 100 ns CG simulation (five structures taken from 0-100 ns trajectory), and melt2 (Figure 6(f)) having large micelles of PE-b-PEG in PE obtained close to the end of 5000 ns simulation (five structures taken from 4900-5000 ns trajectory). Also, five structures (equally spaced in the 400-500 ns trajectory) were taken from each of the two distinct CG PCNC morphologies, viz. PCNC1 (exfoliated) and PCNC2 (tactoid) systems. All four morphologies were converted to AA resolution and properly relaxed before property calculations (procedure described in Methods Section 2.3). The CG models and corresponding backmapped AA structures are shown in Supplementary Information Figure S16.

**Glass Transition Temperature**

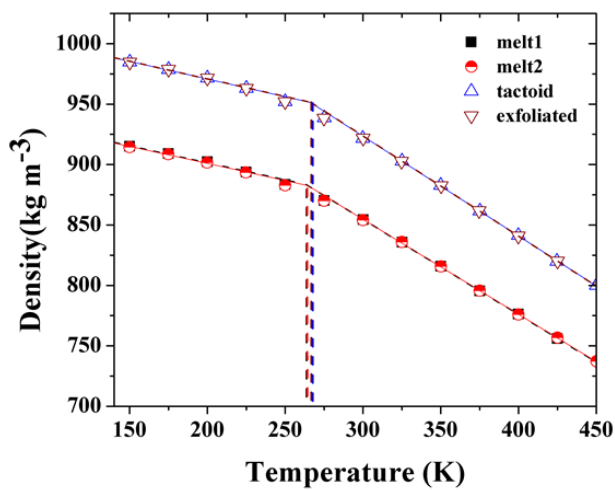

**Figure 9:** Temperature-Density plot of melt and PCNCs. Data obtained from linear cooling (5 K/ns) of backmapped AA structures from 450 K to 140 K. The data points indicate the block average (over 5 ns trajectory, equivalent to 25 K interval) values of density and temperatures). The solid lines are the linear fits in glassy (low temperature) and rubbery regions (high temperature), with the intersection used as as an estimate of $T_g$ (dashed vertical line).

The $\rho$-$T$ curve was generated using data obtained from cooling the system from 450 K to 140 K at a rate of 5 $\frac{K}{ns}$ under *NPT* ensemble. Figure 9 shows the $\rho$-$T$ curve for one run of each of the four systems. As per the free volume theory,[118,119] there is a sharp change in the slope of $\rho$-$T$ curve near the glass transition temperature, $T_g$, as clearly evident in Figure 9. The intersection of two regimes was used as an estimate of $T_g$, giving values in the range 261–263 K for the PE/PE-b-PEG melt systems. This $T_g$ is higher than the experimentally measured $T_g$ of pure PE which lies in the range 190 K - 210 K.[120,121] which can be primarily attributed to the significantly higher cooling rates in MD simulations.[122] The experimental estimate for $T_g$ of PE (molecular weight of $10^5 \frac{g}{mol}$) was also reported to shift from 208 K to 222 K on increasing the cooling rate from 500 $\frac{K}{s}$ to 5000 $\frac{K}{s}$.[123] Here, we used the WLF equation[102] to extrapolate $T_g$ to the experimental cooling rates of 1 $\frac{K}{s}$ and obtained a value of 193.66 K for melt1 and 194.77 K for melt2, in excellent agreement with the experimental values for PE.[120,124,125] $T_g$ of PCNCs is in range 201-202 K, i.e. 10 K higher than that of the melt system (see Supplementary Information Figure S17). This trend is in good agreement with other experimental and simulation studies.[126–128] The addition of weakly interacting nanoparticles is expected to slow polymer dynamics and also affect chain relaxation modes (as evident by a small decrease in $s_c$ seen in Table 5), leading to a small increase in $T_g$.

**Young's Modulus**

The Young's modulus ($E$) was determined from the slope of stress-strain curves at 140 K (significantly below $T_g$) and 200 K (close to $T_g$) (stress-strain curves shown in Supplementary Information Figure S18). The relevant component of the stress tensor was calculated in the limit of slow uniaxial extension or compression. The $E$ of melt2 at 140 K and 200 K was 4.17 GPa and 2.16 GPa, respectively. The near $T_g$ value of $E$ is lower than the experimental values for PE obtained in the range 2.24–3.2 GPa (measured at T $\approx$ 240K).[129] This underestimate from our simulations can be attributed to lower molecular weight of the polymer (PE80) and presence of a low molecular weight plasticizer (PE16-b-PEG3). Young's modulus of the composite systems was higher by only 8-9% at 140 K (4.50 GPa for PCNC1, 4.54 GPa for PCNC2), indicating a lower than expected effect of the filler. Greater interfacial area and strong interactions between the filler and the matrix are known to lead to nanoconfinement effects that facilitate significant enhancement of mechanical properties.[130–132] Young's modulus increased by 2-38% on addition of 2-8% nanoclay in low density polyethylene matrix (LDPE/NC nanoclay composite films prepared by melt extrusion) at room temperature.[133] However, the nanoclay was more hydrophobic (Cloisite 30B), leading to a stronger interaction with PE. For our system, there was a larger increase in $E$ at temperature closer to extrapolated $T_g$ (200 K), with PCNC1 ($E = 2.44$ GPa) and PCNC2 ($E = 2.51$ GPa) values 13-16% higher than the melt value. We expected this difference to further increase at temperatures closer to MD estimated $T_g$ of ~ 267 K. However, the stress-strain curve had a large noise and deviated from linearity at 250 K, precluding estimation of the Young's modulus at this temperature. We conjecture that some local relaxation modes of the chain are unlocked at this temperature, requiring a significantly longer simulation to obtain a relaxed state under applied strain and accurately estimate the stress component.

## 4. Conclusion

We have developed MARTINI FF parameters for TMA-MMT clay particle, wherein the bonded parameters were parmaterized using accurate estimate of bending stiffness and Young's modulus from DFT and all atom MD simulations as constraints while the non-bonded parameters for clay edge beads ($[AlO_2H(OH)]$, $[MgO_2H(OH)]$, $[SiO_2(OH)]$ and $[Si(OH)_2]$) were developed using the dispersive and the polar contributions to the clay cleavage energy. The optimal $P_2$ type for edge beads lies between the alcohol group (P1) and the diol groups (P3 or P4), and thus, corresponds well to the polarity of the edge bead functional groups. The developed CG parameter reproduced AA estimated conformational (error < 3% for $R_g$ and $s_c$) and dynamical (error < 4% for $D$) properties of PE near a periodic TMA-MMT sheet.

The developed CG parameters were used to investigate the long-timescale structural evolution and properties of PE/PE-b-PEG melt and its composite with exfoliated (PCNC1) and tactoid (PCNC2) TMA-MMT assemblies. These simulations allowed capturing of slow clay-induced disruption of block copolymer assembly and its redistribution in the PCNCs, with the compute time for convergence in the evolution of block copolymer partitioning reduced from 178 days for AA to just 2 days. We observed re-orientation of PE-b-PEG chains within the simulation time driven by a favourable interaction of PEG hydroxyl group with either the bridging oxygen atoms or the edge hydroxyl groups of the siloxane surface. This almost complete saturation of the clay surface by the block copolymer improved the solvation of clay particle in the composite, providing an alternative to use of clay particles with large alkyl chain substitutions for improved dispersion. TMA-MMT coated with PE-b-PEG acted as a neutral interaction surface for PE, causing a negligible change in its structural and thermodynamic properties. The primary influence of the nanofiller was shown on the slowdown of dynamics because of steric effects, with a 18% and 12% reduction in the PE diffusion coefficient in PCNC1 (exfoliated clay) and PCNC2 (tactoid clay), respectively. Finally, select properties, such as the glass transition temperature and the Young's modulus, having a high-dependence on details of molecular interaction were determined by backmapping the CG structures to AA resolution. The melt $T_g$ of 193.67 K increased to 201-202 K and the Young's modulus improved by 13-16% for both PCNCs, in agreement with other studies on effect of nanofiller addition. Overall, these observations clearly show the application potential of developed MARITNI CG parameters for rational design of PCNCs for target properties.

ASSOCIATED CONTENT

**Supplementary Information**. The Supporting Information is available in a separate document, which includes the following additional information.

**S1** Determination of coarse-grained interaction parameters, determination of clay cleavage free energy and Martini-LJ well-depth, partitioning of cleavage energy into dispersive and polar components, and additional simulation system details.

**S2** Determination of Young's modulus and bending stiffness of TMA-MMT sheet and additional simulation details.

**S3** Procedure used for obtaining target density and equilibrium properties of the polymer-clay system, additional method details on property calculation, and some key results including RDF, first shell CN, radius of gyration, and conformational entropy of melt and PCNC systems.

**S4** Back mapping procedure for polymer

**S5** AA simulation details of backmapped melt and PCNCs, and additional method details on calculation of Glass Transition Temperature and Young's Modulus

The following files are available free of charge.

SI_Khan2025CG2AA_PCNC.pdf: Supporting information file (file type: PDF file)

GitHub link containing all the structure and simulation parameters for melt, PCNC1, and PCNC2 for both all-atom (AA) and coarse-grained (CG) simulations:

https://github.com/Ankitsheetal/ManuscriptJCTCfiles.git

AUTHOR INFORMATION

**Corresponding Author**

*goelg@chemical.iitd.ac.in

**Author Contributions**

The manuscript was written through contributions of all authors. All authors have given approval to the final version of the manuscript.

**Funding Sources**

P.K. and A.P. thank Ph.D. scholarships by MHRD, Government of India.

ACKNOWLEDGMENT

The authors thank IIT Delhi HPC facility for computational resources.

(35) Huang, M. F.; Yu, J. G.; Ma, X. F. Studies on the Properties of Montmorillonite-Reinforced Thermoplastic Starch Composites. *Polymer (Guildf).* **2004**, *45* (20), 7017–7023. https://doi.org/10.1016/J.POLYMER.2004.07.068.

(36) Mao, C.; Zhu, Y.; Jiang, W. Design of Electrical Conductive Composites: Tuning the Morphology to Improve the Electrical Properties of Graphene Filled Immiscible Polymer Blends. *ACS Appl. Mater. Interfaces* **2012**, *4* (10), 5281–5286. https://doi.org/10.1021/AM301230Q/SUPPL_FILE/AM301230Q_SI_001.PDF.

(37) Calcagno, C. I. W.; Mariani, C. M.; Teixeira, S. R.; Mauler, R. S. The Role of the MMT on the Morphology and Mechanical Properties of the PP/PET Blends. *Compos. Sci. Technol.* **2008**, *68* (10–11), 2193–2200. https://doi.org/10.1016/J.COMPSCITECH.2008.03.012.

(38) Rafiee, R.; Shahzadi, R. Mechanical Properties of Nanoclay and Nanoclay Reinforced Polymers: A Review. *Polym. Compos.* **2019**, *40* (2), 431–445. https://doi.org/10.1002/PC.24725.

(39) Suter, J. L.; Groen, D.; Coveney, P. V. Chemically Specific Multiscale Modeling of Clay–Polymer Nanocomposites Reveals Intercalation Dynamics, Tactoid Self-Assembly and Emergent Materials Properties. *Adv. Mater.* **2015**, *27* (6), 966–984. https://doi.org/10.1002/ADMA.201403361.

(40) Liu, J.; Zhang, L.; Cao, D.; Wang, W. Static, Rheological and Mechanical Properties of Polymer Nanocomposites Studied by Computer Modeling and Simulation. *Phys. Chem. Chem. Phys.* **2009**, *11* (48), 11365–11384. https://doi.org/10.1039/B913511A.

(41) Picard, E.; Gérard, J.-F.; Espuche, E. Reinforcement of the Gas Barrier Properties of Polyethylene and Polyamide Through the Nanocomposite Approach: Key Factors and Limitations. *Oil Gas Sci. Technol. – Rev. d'IFP Energies Nouv.* **2015**, *70* (2), 237–249. https://doi.org/10.2516/ogst/2013145.

(42) Durmuş, A.; Woo, M.; Kaşgöz, A.; Macosko, C. W.; Tsapatsis, M. Intercalated Linear Low Density Polyethylene (LLDPE)/Clay Nanocomposites Prepared with Oxidized Polyethylene as a New Type Compatibilizer: Structural, Mechanical and Barrier Properties. *Eur. Polym. J.* **2007**, *43* (9), 3737–3749. https://doi.org/10.1016/j.eurpolymj.2007.06.019.

(43) Ganesan, V.; Jayaraman, A. Theory and Simulation Studies of Effective Interactions, Phase

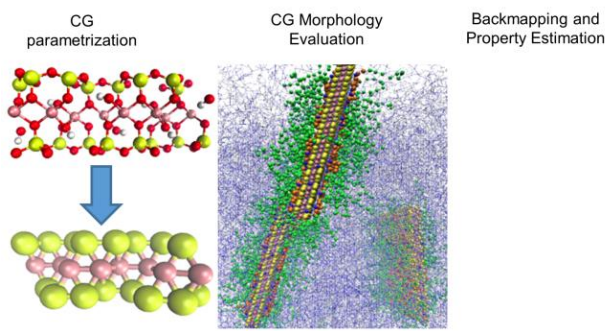

CG parametrization · CG Morphology Evaluation · Backmapping and Property Estimation

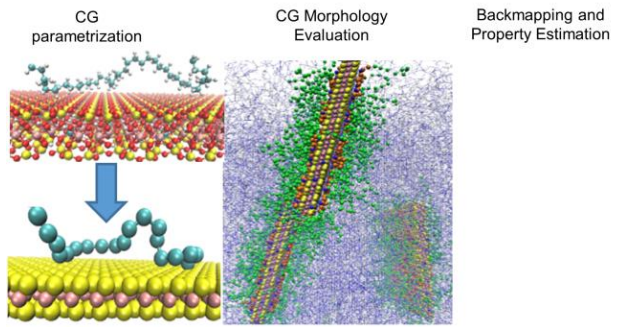

CG Morphology · AA Morphology · Property Estimation

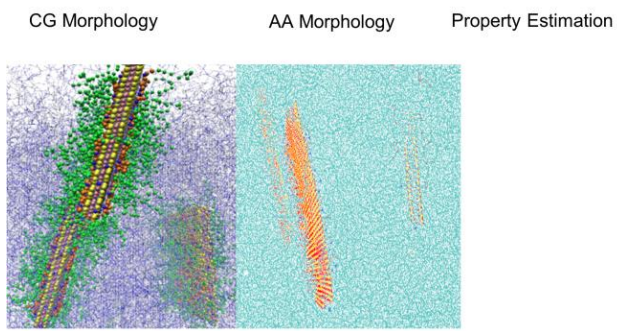